\documentclass[%
 aip,
 amsmath,amssymb,
 reprint,%
]{revtex4-1}
\usepackage{gensymb}
\usepackage{comment}
\usepackage{xcolor}
\usepackage{graphicx}% Include figure files
\usepackage{dcolumn}% Align table columns on decimal point
\usepackage{bm}% bold math
\usepackage{amsmath }
\usepackage[utf8]{inputenc}
\usepackage[T1]{fontenc}
\usepackage{mathptmx}
\usepackage{etoolbox}
\usepackage{titlesec}
\titlespacing\section{0pt}{6pt plus 2pt minus 2pt}{6pt plus 2pt minus 4pt}
\titlespacing\subsection{0pt}{6pt plus 2pt minus 2pt}{6pt plus 2pt minus 2pt}
\makeatletter
\def\@email#1#2{%
 \endgroup
 \patchcmd{\titleblock@produce}
  {\frontmatter@RRAPformat}
  {\frontmatter@RRAPformat{\produce@RRAP{*#1\href{mailto:#2}{#2}}}\frontmatter@RRAPformat}
  {}{}
}%
\makeatother
\begin{document}

\preprint{AIP/123-QED}

\title{Collimated versatile atomic beam source with alkali dispensers}
% Force line breaks with \\
\affiliation{ 
School of Physics, Georgia Institute of Technology, 837 State Street, Atlanta, Georgia 30332, USA}%

\author{Bochao Wei}
\email{bwei39@gatech.edu, lichao@gatech.edu}

\author{Alexandra Crawford}

\author{Yorick Andeweg}
 \altaffiliation{Currently at Physics Department, CU Boulder}
\author{Linzhao Zhuo}
\author{Chao Li}%
\author{Chandra Raman}

\date{\today}% It is always \today, today,

\begin{abstract}
Alkali metal dispensers have become an indispensable tool in the production of atomic vapors for magnetometry, alkali vapor cell clocks, and laser cooling experiments.  A primary advantage of these dispensers is that they contain alkali metal in an inert form that can be exposed to air without hazard.  However, their high temperature of operation (> 600 $\degree$C) is undesirable for many applications, as it shifts the atomic speed distribution to higher values and presents a radiative heat source that can raise the temperature of its surroundings.  For this reason, dispensers are typically not used in line-of-sight applications such as atomic beam generation.  In this work, we present an integrated rubidium dispenser collimating device with a thickness of only 2 mm that produces a beam of atoms traveling primarily in the forward direction. We find that the collimator plate serves to both shield the dispenser’s radiation as well as to moderate the velocity of the atomic beam so that the measured longitudinal speed distribution is comparable to that of an ordinary alkali oven at only a slightly elevated temperature of 200$\degree$C. To confirm our 
theory, we also constructed another compact apparatus consisting of a dispenser and a silicon collimator and the measurements support our conclusion. 
Our integrated dispenser collimator will particularly be useful in integrated photonics and cavity QED on chip, where a localized, directed source of Rb vapor in small quantities is needed.\end{abstract}

\maketitle

 Alkali atoms are a key resource for numerous emerging chip-scale quantum technologies\cite{keil2016fifteen,kitching2018chip}.  With the ease of batch fabrication methods, chip-based atomic devices such as vapor cell atomic clocks and magnetometers\cite{maurice2017microfabricated,griffith2010femtotesla,sebbag2021demonstration, sheng2017microfabricated},  quantum gravimeter\cite{abend2016atom}, and Bose-Einstein condensates on chip\cite{colombe2007strong,berrada2013integrated,treutlein2007bose,jo2007long} have become increasingly attractive for realizing novel quantum device architectures that can be more widely disseminated.  As this miniaturization trend continues, there is a growing need to integrate small-scale alkali atom sources directly on chip or within a small volume vacuum cell.  
Alkali metal dispensers\cite{gettersusa} are small in size and can be handled in ambient air.
 However, the high temperature needed to initiate the reactions that produce alkali vapor (550-850 $^{\circ}$C) creates significant challenges for integration into a chip environment, as they present a considerable source of radiative heat that can influence other chip components.  Moreover, at such elevated temperatures, the emitted atomic flux from a dispenser has a substantial longitudinal velocity and must first thermalize with room temperature surfaces in order to be captured in a magneto-optical trap\cite{roach2004novel}.  For this reason, dispenser activation and subsequent utilization of the rubidium vapor are often performed in two separate steps.

In this work, we demonstrate a compact technique for the generation of directed atomic beams from an alkali dispenser, which can be useful for miniature applications demanding line-of-sight to the alkali source.  For such applications, the broad angular distribution of alkali vapor emitted from a bare dispenser is unacceptable since it can degrade the signal-to-noise ratio as well as contaminate nearby electronic or photonic components. In our approach, laser micromachined holes in a collimator plate deliver atoms primarily in the forward direction.  We also find that as the beam is generated, atoms rapidly thermalize with the collimator to a considerably lower temperature, as determined by measurements of the longitudinal velocity. Combined with its small size, this collimated source can be easily packaged close to other chip-scale components. Our device will find application in targeted delivery of neutral atoms to microscope volumes on chip, including on-chip cavities or nanophotonic devices for cavity QED\cite{dorche2020high,alaeian2020cavity}.  
%With the advent of nanotechnology, integrated photonics, MEMS and chip-scale atomic clocks, there is a need for small scale alkali atom source and integrating the source directly on chip near the locations of specific on-chip elements. Microfabricated alkali vapor cells [ref] are frequently used but they suffer from broad Doppler shift and collisions [ref]. Furthermore, for some applications that require atoms flying close to the proximity of nanophotonic structures for interaction, it is preferred to have a controlled way of delivering alkali atoms.

\begin{figure*}[htbp]
\centering
\includegraphics[trim = 0 0 0 0, width=0.8\linewidth]{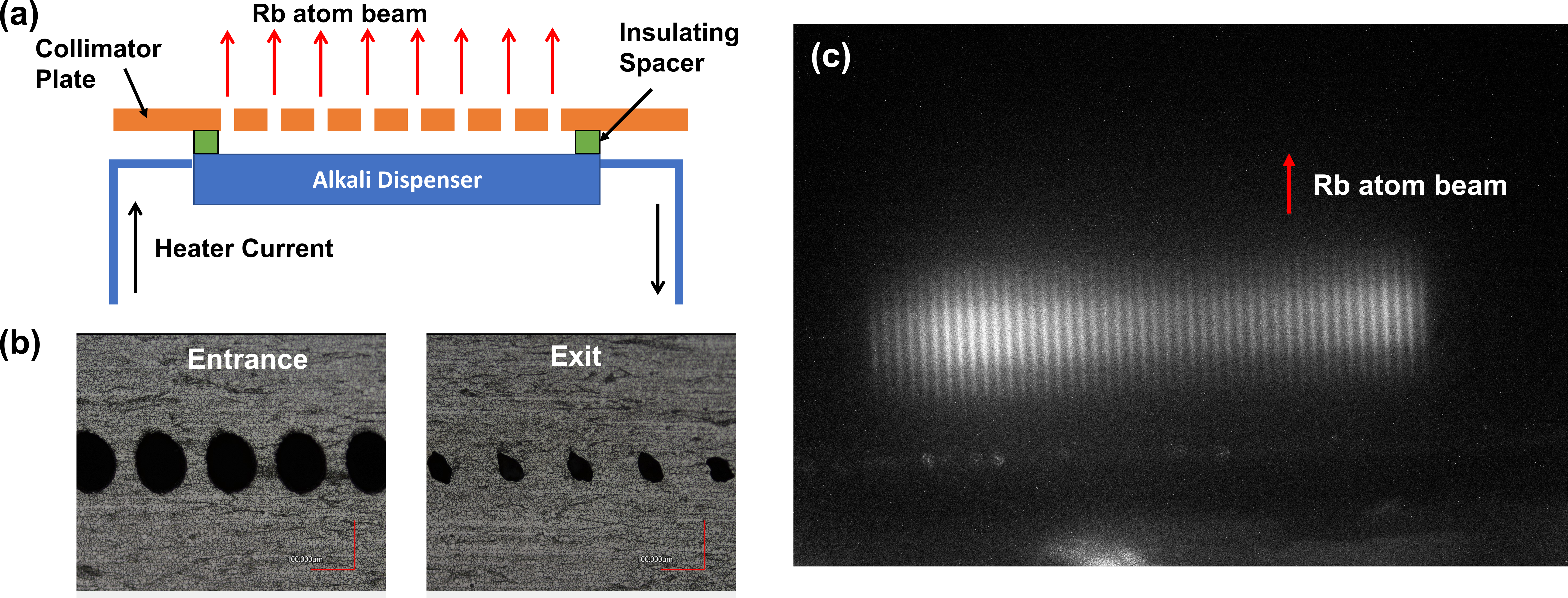}
\caption{(a) Overview of the atomic beam apparatus (b) the detailed view of the entrance and exit channel shape, the length of the scale bar is 100 $\mu$m. The entrance face towards the dispenser. (c)The CCD camera image of the atomic beams fluorescence. The laser beam enters from the right and the atomic beams are traveling towards the top.}
\label{fig:1} 
\end{figure*}

We used SAES alkali metal dispensers as the atomic source. It holds a mixture of rubidium chromate and a reducing agent within a metal container, which has a trapezoidal cross-section with a small slit to allow alkali metal vapor to exit. The active area of the dispenser is around 6 mm $\times$1 mm. The apparatus is shown schematically in Fig.~\ref{fig:1}(a). We use a 600 $\mu$m thick stainless steel plate with micro-channels to collimate the emitted rubidium atoms. This collimator plate is fabricated using femtosecond laser micromachining technique (OPTEC WS-Flex USP).  There are 46 channels in the center of the plate with a spacing of 160 $\mu$m that fully covers the active length of the dispenser. The aspect ratio of the channels is around 6:1. As a result of this aspect ratio and the nature of laser micro-machining, the fabricated channels have a varying diameter across the plate. As shown in Fig.~\ref{fig:1}(b), the entrance diameter is around 115 $\mu$m while the exit is an ellipse with $2a\approx45\ \mu$m and $2b\approx70\ \mu$m. The plate is positioned with the entrance facing the dispenser to achieve better collimation. The overall device dimensions are around 18 mm$\times$ 2 mm$\times$ 2 mm, and could be further reduced.

A high-temperature ceramic adhesive (PELCO) is applied between the dispenser and the collimator to hold them together while serving as an insulating spacer. This adhesive provides both ultra-low electrical and thermal conductivity so that the current mostly runs through the dispenser and creates a temperature difference between the dispenser and the collimator. The adhesive also seals the space between the dispenser and the collimator to avoid vapor leakage. Although this procedure requires handling the dispenser in ambient air, we have noticed that the dispensers are not noticeably degraded.

The device was activated by running an 8 A current for around 10 minutes. After activation, the current was lowered to around 6 A and a relatively steady flux of rubidium atoms was produced. The operating current is greater than the bare dispenser because of the small exit area ($\sim$0.12 mm$^2$) and the extra heat load from the collimator plate and ceramic adhesive.

\begin{figure*}[htbp]
\centering
\includegraphics[width=0.8\linewidth]{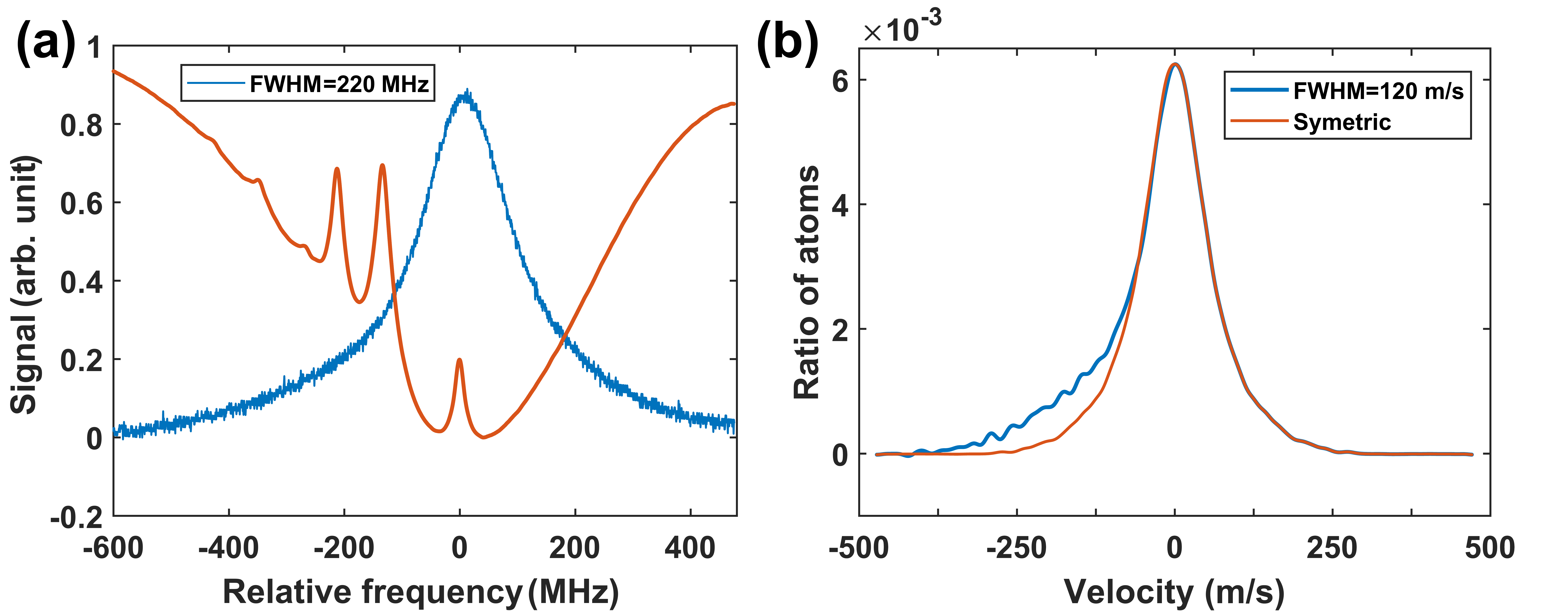}
\caption{(a) Measured fluorescence spectrum with the saturated spectrum.(b) The speed distribution after deconvolution. The red curve is the symmetric distribution from a ideal two level system. The discrepancy around  negative 200 m/s results from the influence of F=2 to F'=2, 1 transitions. }
\label{fig:spectrum} 
\end{figure*}

The device was put inside a cubic vacuum chamber that reached a pressure around 4$\times10^{-7}$ torr. Electrical inline connectors and two feedthrough pins were used to connect the dispenser to the outside of the chamber. A laser beam was sent perpendicular to the atomic beam direction. Laser spectroscopy at the rubidium D2 line is used to measure the transverse speed distribution. A current of 6.5 A was run through the dispenser and the fluorescence was collected by using a 2 inches lens set with a numerical aperture around 0.24 and a silicon photodetector (DET100A2). The photocurrent passes through a current preamplifier (Model 1211 DL Instruments) with a gain of $10^9$ V/A. Figure \ref{fig:1}(c) is the image of the beam fluorescence with our laser locked to the rubidium D2 $F=2$ to $F'=3$ transition. All 46 channels are clearly visible. 

The atomic beam fluorescence spectrum is shown in Fig. \ref{fig:spectrum} (a). The full width at half maximum (FWHM) is 220 MHz. The spectrum is a convolution between the scattering rate $R_{sc}$ and the atoms' transverse Doppler distribution $n(v_t)$. Using the deconvolution technique \cite{li2020robust}, and given our laser intensity $I=2$ mW with beam size $w=0.586$ mm, we can deduce the transverse speed distribution as shown in Fig.~\ref{fig:spectrum} (b).
The FWHM for the transverse speed distribution is around 120 m/s. The deconvolution process here only considered a single $F=2$ to $F'=3$ hyperfine level, while the $F=2$ to $F'=2,1$ transitions contribute to the asymmetric shape around $-200$ m/s \cite{li2020robust}.

\begin{figure*}[ht]
\centering
\includegraphics[width=0.8\linewidth]{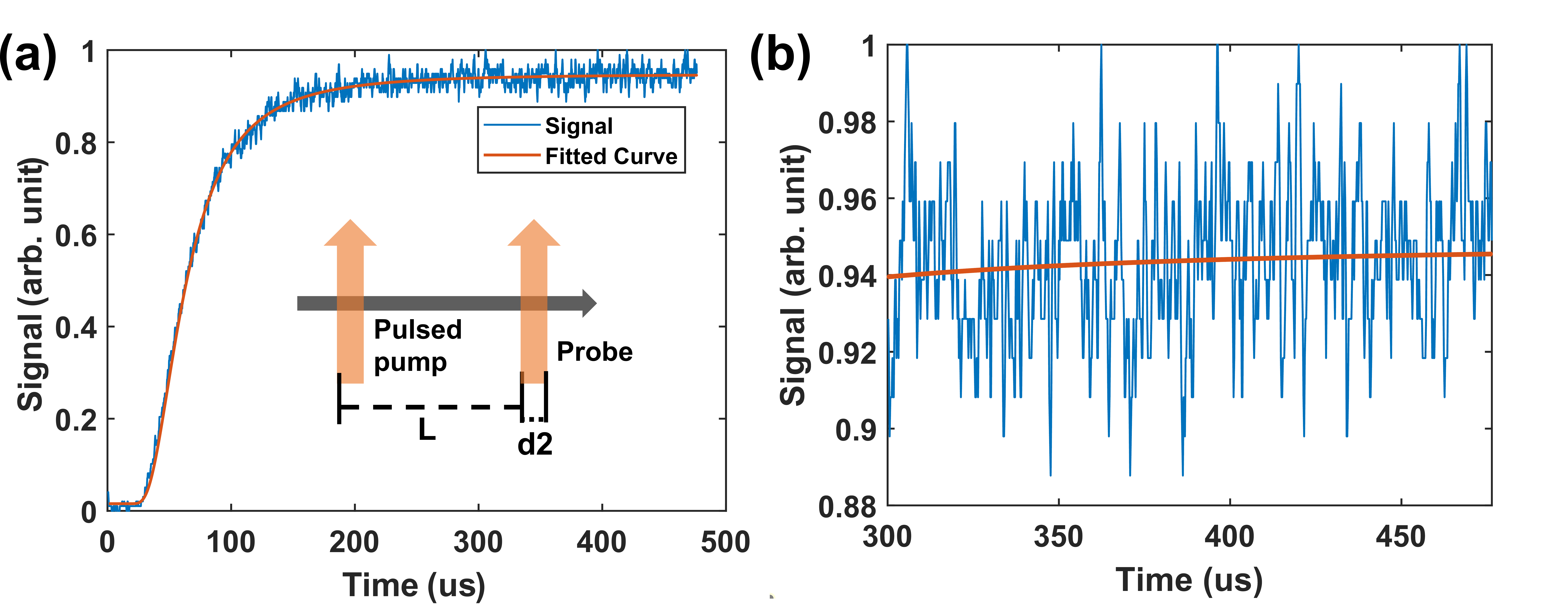}
\caption{(a) The recover signal of the time of flight measurement. Time zero is when the pump laser pulse is switched off. The fitted curve is assuming a Maxwell distribution with $u=302$ m/s. The inset is the diagram for the time of flight set up. (b) Zoomed-in plot for the signal of slow atoms between 44.5 m/s and 70m/s. The red line is the theory curve.}
\label{fig:tof} 
\end{figure*}

The longitudinal velocity distribution is measured by using a modified time of flight technique\cite{molenaar1997diagnostic}. We found this method to be more convenient than  measuring the longitudinal Doppler shift spectrally since the latter approach would be easily affected by the transverse velocity distribution, power broadening, and other hyperfine transitions. In our time-of-flight technique, a locked pumping laser is tuned to the Rb D2 $F=2$ to $F'=2$ transition that selects atoms with nearly zero transverse velocity, pumping them into the dark hyperfine ground state $F = 1$ as shown in Fig. \ref{fig:tof}(a) inset. A probe beam locked to the Rb D2 $F=2$ to $F'=3$ transition is located downstream from the pump. The fluorescence from the probe beam is collected by a micro-photomultiplier tube (Hamamatsu H12403-20). The laser beam separation $L=20$ mm, while the diameters of both beams are $d1=d2=1.2$ mm. The dispenser current was increased to 7 A to increase the signal-to-noise ratio.
While the pump beam is on, all atoms enter the dark state and the probe fluorescence will be zero. By switching off the pump, atoms in the bright state $F = 2$ will enter the probe region with a time of arrival that depends on their velocity. Thus following the switch-off, the detector records a time-dependent fluorescence signal $S(t)$ that starts from zero and gradually reaches a steady state.  Figure \ref{fig:tof}(a) shows the data. The fluorescence signal in the steady state will depend on the density of atoms in the probe region. Assuming a Maxwell Boltzmann longitudinal speed distribution\cite{ramsey1985molecular}, we can write our signal as:
\begin{equation}
 S(v,t) = \begin{cases} 
          0 & \ vt\leq L \\
          c\cdot v^2e^{-\frac{-v^2}{u^2}}\cdot (vt-L) & L\leq vt\leq L+d2 \\
          c\cdot v^2e^{-\frac{-v^2}{u^2}}\cdot d2 & vt\geq L+d2
       \end{cases}
\end{equation}
In which $L$ is the distance between the pump beam and the probe beam, $d2$ is the diameter of the probe beam, $c$ is the amplitude factor that does not depend on $v$, and $u=\sqrt{2kT/m}$ is the most probable speed. Then, our total signal at time $t$ including all velocity groups is:

\begin{equation}
\begin{split} 
\label{eqn:signal}
    &S(t)=\int_{L/t}^{\infty} S(v,t)dv\\
    &=c(\int_{\frac{L}{t}}^{L+d2/t}v^2 e^{\frac{-v^2}{u^2}}\cdot(vt-L)dv+\int_{\frac{L+d2}{t}}^{\infty}v^2 e^{\frac{-v^2}{u^2}}\cdot d_2 dv)
\end{split}
\end{equation}
This formula is fitted to our signal as shown in Fig. \ref{fig:tof}, yielding a peak velocity of $u=302$ m/s, corresponding to a Maxwell distribution with a temperature of around 204 \degree C. This temperature is significantly lower than the expected operating temperature of the dispenser (> 600\degree C), indicating that the atoms thermalized with the colder surfaces of the collimator plate before exiting the device. Thus it produces slower atoms on average, compared with the bare dispenser. These findings suggest that direct line-of-sight laser cooling might be possible using this integrated dispenser collimator, thus avoiding contamination of the vacuum chamber or miniature cell.

To further quantify the slow atoms, by using $d_2<<L$ and $v_i=\frac{L+d_2}{t_i}$ in equation (\ref{eqn:signal}), we can get 
\begin{equation}
\label{equ:diff}
    S(t_1)-S(t_2)\sim c\int_{v_1}^{v_2}v^{2}e^{-\frac{v^2}{u^2}}dv
\end{equation}

This means that the difference of $S(t)$ between time $t_1$ and $t_2$ is proportional to the density of atoms with velocity $v_1$ and $v_2$. Figure \ref{fig:tof} (b) shows the zoomed-in plot between $t_1=476\ \mu s, t_2=303\ \mu s$ which corresponding to $v_1=44.5$ m/s and $v_2=70$ m/s. By using the fitted $S(t)$ and normalization $S(\infty)-S(0)=1$, the atom density within $v_1=44.5$ m/s and $v_2=70$ m/s is calculated to be 0.61\% of the entire atom density. By comparison, the theoretical ratio for a Maxwell distribution is 0.67\%, which is very close to the measured value.  To further confirm these observations, we implemented the likelihood ratio test between the null hypothesis and the alternative hypothesis. The null hypothesis is that there are no slow atoms in that velocity range, thus $S(t)$ should be flat and the observed signal is just Gaussian noise. The alternative hypothesis is that the theoretical equation (\ref{equ:diff}) is valid for slow atoms and the observed signal is $S(t)$ plus Gaussian noise. The likelihood ratio of the alternative hypothesis to the null hypothesis is calculated to be $1.1\times 10^4$. Thus the signal shows strong evidence for the existence of slow atoms in a thermal beam. These slow atoms are easier to control and can be captured by a MOT downstream or might be used directly for on-chip applications.

\begin{figure}[t]
 \includegraphics[width=0.5\textwidth]{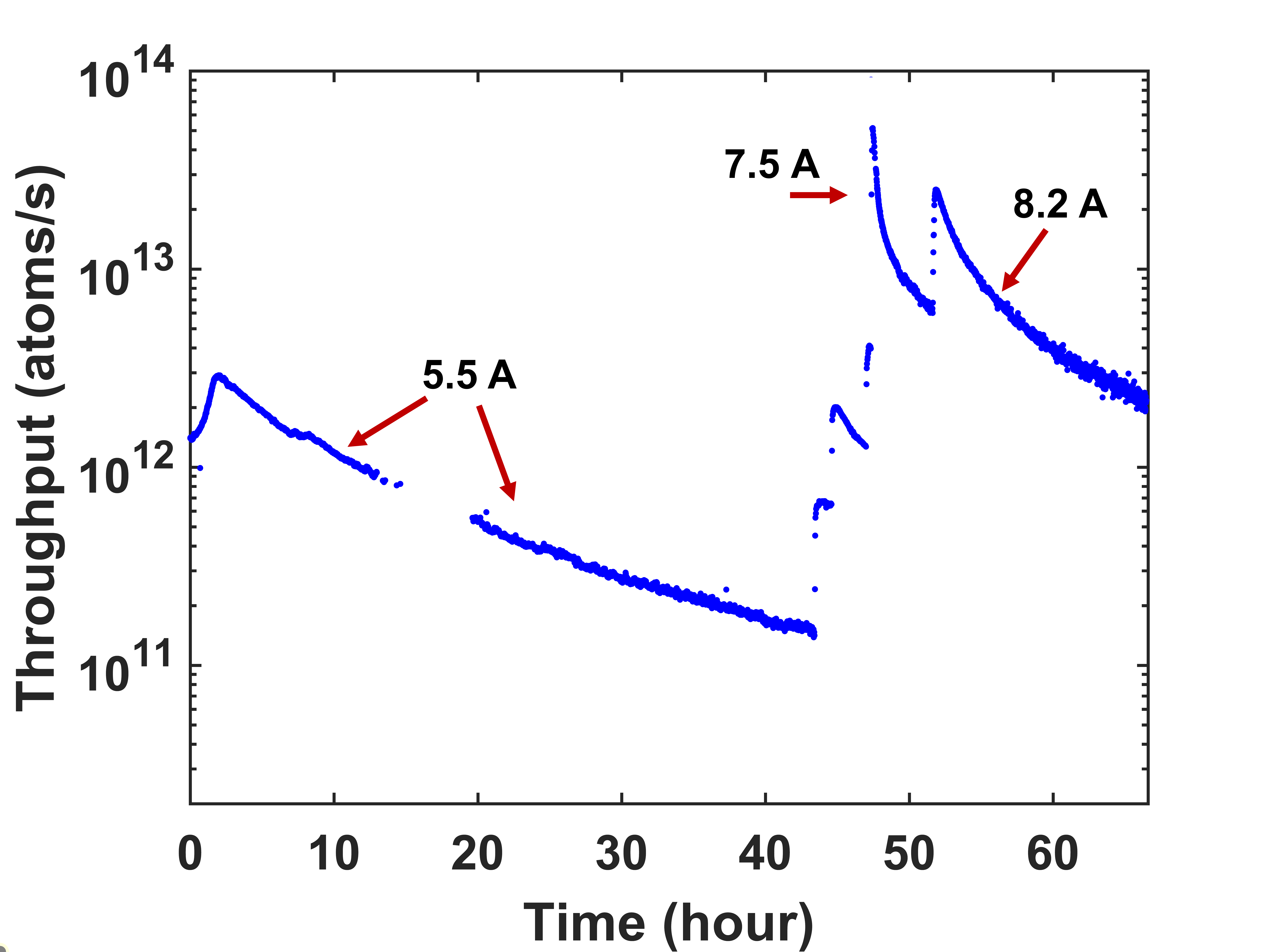}
 \caption{Throughput vs time curves of the integrated dispenser collimator. 
 %current was run through the dispenser in intervals of $\sim$0.5 A ranging from 5.5 A to 12.8 A and the
  Curves with a current of 5.5 A, 7.5 A and 8.2 A are labeled and currents higher than 8.2 A exhibited similar dynamics. Overnight, the laser scan region drifted out of the Rb87 transition and caused a gap in the 5.5 A data.}
 \label{fig:lifetime} 
 \end{figure}

The long-term performance of this device was characterized by monitoring the flux at variable supply currents. The initial current was 5.5 A and it was increased in steps of $\sim$0.5 A to 8.2 A. %accelerate the lifetime measurement. 
The total throughput of the device at different currents are shown in Fig. \ref{fig:lifetime}. 
%Two typical flux curves at 5.5 A and 8.2 A are indicated in Fig. \ref{fig:lifetime}. 
With an elevated current through the dispenser, the flux rapidly increased before slowly decreasing on a time scale of hours. This behavior is typical for alkali dispensers, according to its spec sheet\cite{gettersusa}. A feedback loop may be integrated into the system in the future to produce a constant flux. After 8.2 A, the input current to the dispenser was gradually increased to 12.8 A to test when the device will fail. After around 20 minutes at 12.8 A, the collimator plate detached because of the thermal expansion mismatch between the ceramic adhesive and metal surfaces. This indicates the current limit of this device should be around 12 A.

The total test last around 75 hours. By integrating the area under the curves, we calculated that the rubidium emitted during this test is around 0.18 mg. Our device lifetime is greatly increased due to the collimator, which blocks and saves the off-axis atoms while maintaining the on-axis flux. Applications requiring higher collimation can use a collimator with a higher aspect ratio and would have an even longer lifetime.

\begin{figure*}[htbp]
    \centering
    \includegraphics[width=0.77\textwidth]{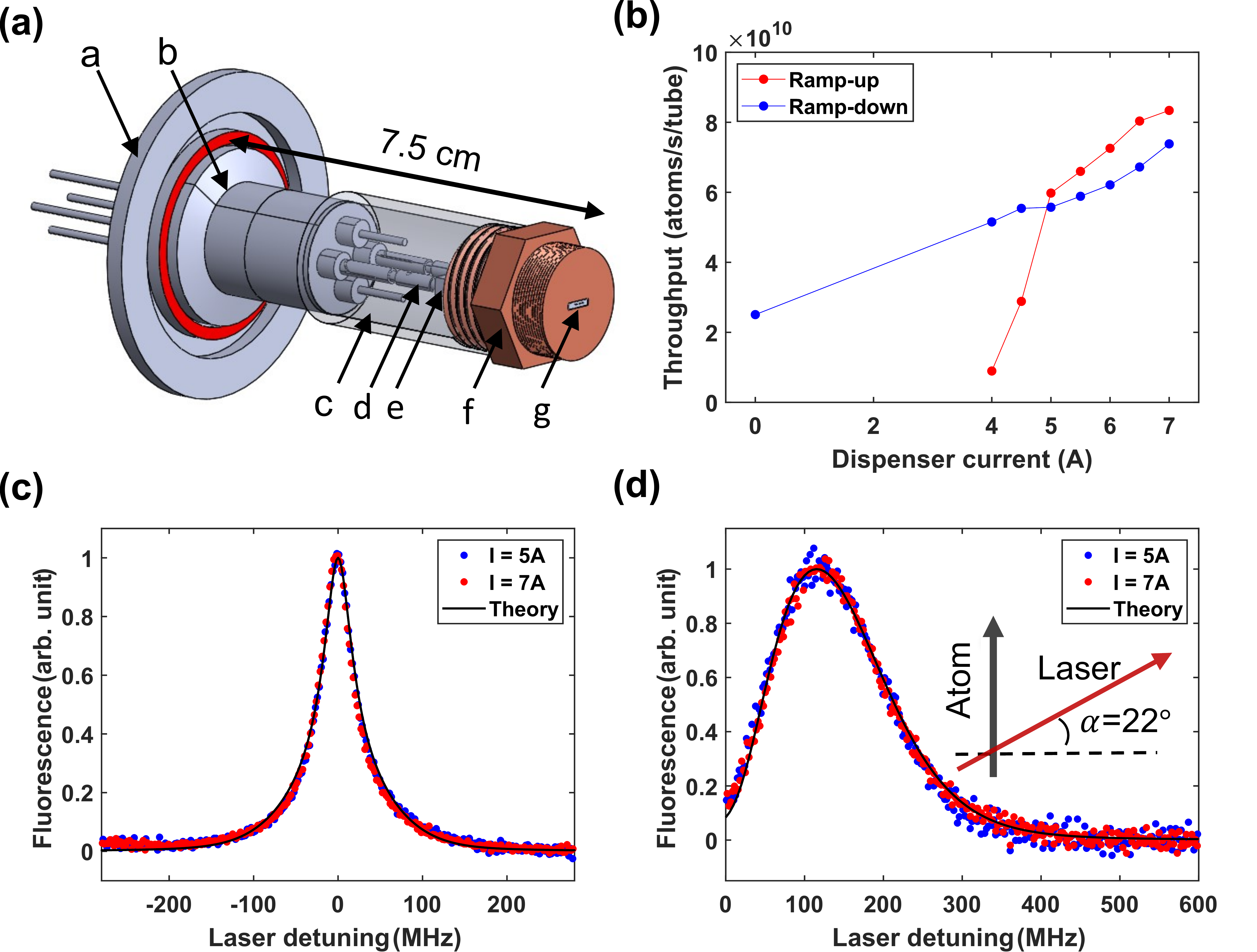}
    \caption{(a) CAD drawing of the atomic beam oven assembly. a. KF40 bored blank. b. power feedthrough. c. Stainless steel tube. d. Crimp connectors. e. Dispenser. f. CNC machined holder with an EDM slit at the center for hosting the silicon collimator. g. An ordinary silicon collimator. (b) Total throughput of the atomic beam oven vs. dispenser current. (c) Measured fluorescence spectra. The FWHM for 5 A and 7 A are both 50MHz. (d) Measured fluorescence spectra when the laser beam is tilted around $\alpha=22\degree$  as shown in the inset. This angle is limited by the geometry of our vacuum chamber.}
    \label{fig:oven and data}
\end{figure*}
\bigskip

Our observations so far indicate that the hot atoms from the dispenser rapidly thermalize with the collimator plate with a transit distance of only 600 $\mu$m.  As a result, the temperature of the atoms is lowered by a factor of 3 in comparison with the bare dispenser. To explore this effect in more depth, we also built a conventional dispenser oven design in a stainless steel tube of 3/4 inch diameter. We have found that the conventional design performed similarly to the integrated design, which supports our theory.

This dispenser oven design is presented in Fig.~\ref{fig:oven and data}(a). A standard center-bored KF40 flange is laser-welded to another electrical KF feedthrough. The laser welding region is highlighted in red. Pins for electrical connections are sleeved into a stainless steel tube 
%(shown as semi-transparent for viewing components inside the oven cavity) 
with female NPT threads. Another customized copper holder with mating male threads hosting our straight silicon collimator is attached. Here we used the silicon collimator design from our previous work which can provide 30:1 collimation with a size of only 3 mm$\times$5 mm \cite{li2019cascaded}. 

A vacuum-compatible polyimide flexible heater is attached to the outside of the stainless steel tube. A thermistor is glued between part b and part c to serve as the temperature input for a PID controller maintaining the oven body temperature at a set point of 100 $\degree$C. No heater is needed for the copper head whose temperature was measured to be 121$\sim$125 \degree C.

The throughput measurement with different currents
are shown in Fig.~\ref{fig:oven and data}(b). Fifteen spectra were taken sequentially as we first ramped up, then ramped down the current of the dispenser. The time step was 5 $\sim$ 7 minutes. The throughput curves for the ramp-up and ramp-down processes do not lie on top of one another. This implies that the time step we used was not long enough for the oven to reach equilibrium. 

Figure~\ref{fig:oven and data}(c) shows two typical fluorescence spectra taken at 5 A and 7 A. The FWHM at a current of 5 A was 50 MHz, smaller than the 220 MHz value obtained for the integrated design.  This is consistent with a factor of 5 larger aspect ratio used -- 30:1 rather than 6:1 for the stainless steel collimator. The FWHM for all spectra taken within 5 A to 7 A are in the range of 45 $\sim$ 50 MHz, which are roughly the same as previously reported by our group for a 100 \degree C vapor source based on pure Rb metal \cite{li2019cascaded}. 

To estimate the longitudinal speed distribution, the laser beam is tilted by $\alpha=$22$\degree$ with respect to the orthogonal direction of the atomic beam (Fig~\ref{fig:oven and data}(d)). In this case, the fluorescence signal is also sensitive to the longitudinal velocity components. If we assume the longitudinal velocity follows a Maxwell Boltzmann distribution, we can write the fluorescence signal as \cite{beijerinck1975velocity,li2019cascaded}:
\begin{equation}\label{eq:fluorescence_theory}
    V(\delta)\displaystyle \propto \int\limits_\theta \int\limits_v R_{sc}(s,\Delta)F(v) \kappa f(\theta) \mathrm{d}v \mathrm{d}\theta.
\end{equation}
Here, $R_{sc}$ is the scattering rate function that depends on the saturation parameter $s$ and $\Delta=\delta-kv\sin(\theta+\alpha)$, $\delta$ is the laser detuning, v is the atom velocity, $\theta$ is the polar angle, and $\alpha$ is the tilted angle 22$\degree$, $F(v)$ is the Maxwell-Boltzmann distribution with temperature as the fitting parameter, and $\kappa f(\theta)$ is the angular distribution function associated with this silicon collimator which has been well studied\cite{li2020robust}.
By fitting our theoretical curve to the experimental data in Fig~\ref{fig:oven and data}(c) and (d), we found that the temperature of $F(v)$ is around 110 $\degree$C and the most probable speed is around 270 m/s. The fitted temperature is very close to the average of the measured oven body temperature and copper head temperature. 
%These measurements independently confirm that the hot atoms from the dispenser thermalize with the collimator wall and oven wall first before emerging from the collimator. 
This design allow us to monitor the temperature of the oven and independently confirm that the hot atoms from the dispenser thermalize with the collimator wall and oven wall first before emerging from the collimator.

We have demonstrated a versatile and compact approach to creating atomic beams using a collimator-integrated dispenser and compared the results with a more conventional oven design.  In the new design, the collimation plate is integrated onto the dispenser itself, which serves the dual purpose of shielding the environment from the high dispenser operation temperature as well as reducing the effective temperature of the atomic beam.  Our results show that atomic beam technology can be considerably miniaturized without compromising any of its useful properties, which will extend its applications to small-scale quantum devices.  For example, even further integration might be achieved in the future by incorporating dispenser material directly into the collimation plate.
\bigskip

We thank Scott Elliott, Nathan Mauldin, and Frank Murdock at the Montgomery Machining Mall for manufacturing oven test assemblies. We thank Richard Shafer at the IEN Laser Lab for guidance on the Optec femtosecond laser micromachining system. We  acknowledge the support from National Science Foundation (2011478), Office of Naval Research (N00014-20-1-2429). This work was supported by Air Force Office of Scientific Research (FA9550-19-1-0228). 

\section*{AUTHOR DECLARATIONS}
\subsection*{Conflict of Interest}
The authors have no conflicts to disclose.
\section*{DATA AVAILABILITY}
The data that support the findings of this study are available from the corresponding author upon reasonable request.
\section*{REFERENCES}
\bibliography{references}% Produces the 

\end{document}